\documentclass[
 reprint,
 amsmath,amssymb,
 aps
]{revtex4-1}

\usepackage{graphicx}
\usepackage{dcolumn}
\usepackage{bm}
\usepackage{hyperref}
\usepackage[subrefformat=parens]{subcaption}
\usepackage{color}

\begin{document}

\preprint{HUPD-2403}

\title{Non-thermal particle production in  Einstein-Cartan gravity with modified Holst term and non-minimal couplings}

\author{Tomohiro Inagaki}
\email{inagaki@hiroshima-u.ac.jp}
\affiliation{Graduate School of Advanced Science and Engineering, Hiroshima University,
Higashi-Hiroshima~739-8526,~Japan}
\affiliation{Information Media Center, Hiroshima University, Higashi-Hiroshima 739-8521, Japan}
\affiliation{Core of Research for the Energetic Universe, Hiroshima University, Higashi-Hiroshima 739-8526, Japan}
\author{Naoki~Yoshioka}
\email{na-yoshioka@hiroshima-u.ac.jp}
\affiliation{Graduate School of Advanced Science and Engineering, Hiroshima University,
Higashi-Hiroshima~739-8526,~Japan}

\date{\today}

\begin{abstract}
 Non-thermal fermionic particle production is investigated in Einstein-Cartan modified gravity with a modified Holst term and non-minimal couplings between the spin connection and a fermion. By using the auxiliary field method, the theory is rewritten into a pseudoscalar-tensor theory with Einstein-Hilbert action and canonical kinetic and potential terms for a pseudoscalar field.
  The introduced field is called Einstein-Cartan pseudoscalaron.
  If the potential energy of the Einstein-Cartan pseudoscalaron dominates the energy density of the early universe, it causes inflationary expansion.
  After the end of inflation, the pseudoscalaron develops a large value and the non-minimal couplings destabilize the vacuum.
  Evaluating the non-thermal fermionic particle production process, we obtain the mass and the helicity dependences of the produced particle number density.
  We show the model parameters to enhance the preheating and reheating processes.
\end{abstract}

\maketitle

\section{Introduction}
The extension of general relativity (GR) from a geometrical perspective is one of the candidates for solving cosmological problems.
In $F(R)$ theories, the gravitational action is replaced by a function of the curvature, $R\rightarrow F(R)$, and it has been shown to describe well various cosmological phenomena \cite{Starobinsky:1980te,Sotiriou:2008rp, Nojiri:2010wj, Nojiri:2017ncd}.
For example, the idea explains the inflationary expansion of the universe, and the gravitational interaction produces particles necessary to reheat the universe \cite{Kofman:1994rk,Kofman:1997yn}. 
Through the conformal transformation, the additional degree of freedom in the modified action can be represented as a dynamical scalar field that plays a role in the inflaton.
After the end of the inflation, the interaction between the inflaton and the matter converts the inflaton energy into the matter and reheats the universe \cite{Motohashi:2012tt}.
Whether the dominant reheating process is perturbative or non-perturbative particle production
depends on the structure of the interaction.

Recently \cite{Pradisi:2022nmh,Barker:2024dhb}, in the metric-affine gravity \cite{Hehl:1994ue} where the metric $g_{\mu\nu}$ and the affine connection ${\Gamma^{\mu}}_{\nu\rho}$ are independent variables, it is shown that a dynamical pseudoscalaron from the modification of geometrical quantity are obtained due to the existence of the Holst term $\epsilon R=\epsilon^{\mu\nu\rho\sigma}R_{\mu\nu\rho\sigma}/\sqrt[]{-g}$ \cite{Nelson:1980ph, Hojman:1980kv, Holst:1995pc} which consists of the antisymmetric part of the affine connection(torsion) ${\Gamma^{\mu}}_{\nu\rho}-{\Gamma^{\mu}}_{\rho\nu}$.
Consequently, that pseudoscalaron can be used as the inflaton \cite{Salvio:2022suk, Gialamas:2022xtt}.
Thus, even in Einstein-Cartan (EC) gravity (metric-compatible metric-affine gravity) \cite{Cartan:1924yea, Kibble:1961ba,Sciama:1964wt,Hehl:1971qi,Hehl:1976kj},
the inflation can be realized through the pseudoscalar field \cite{DiMarco:2023ncs, He:2024wqv} since there exists the torsion.
The interaction between torsion and matter fields can lead to non-thermal particle production. Several relevant studies exist in this area. For instance, the preheating process has been investigated within Einstein-Cartan gravity incorporating the Nieh-Yan topological invariant \cite{Piani:2023aof}. It has also been noted that the Holst term can either suppress or enhance the rate of vacuum decay \cite{Gialamas:2023emn}.

By introducing an auxiliary pseudoscalar field, 
the Euler-Lagrange equation about affine connection gives the algebra equation regarding the torsion.
Since, in this case, the torsion is represented by the metric, the auxiliary pseudoscalar field, and the matter coupling to the affine connection,
integrating out the torsion yields the effective metric theory as the Palatini $f(R)$ gravity \cite{Sotiriou:2008rp}.
In this effective metric theory, one can get the pseudoscalaron and the interaction between the pseudoscalaron and the matter therefore can realize the inflation and the reheating of the universe.

In this paper, we discuss the non-thermal particle production in EC gravity with the modified Holst term.
Since fermions naturally couple to the torsion in EC gravity, they are considered as matter fields.
Additionally, the natural extension of the kinetic term of the fermion \cite{Freidel:2005sn} is considered.
To discuss the non-thermal fermionic particle production,  we follow a way of previous work about fermionic preheating \cite{Greene:1998nh, Greene:2000ew, Giudice:1999fb, Peloso:2000hy, Herring:2020cah}.
In the discussion, it is assumed that the fermionic field operator is composed of (anti) particles.
By this assumption, one of the new parameters $\alpha$ must vanish.
Numerical calculations eventually reveal that particle production occurs when the value of $\beta$ is large, and its behavior depends on the mass and helicity of the particle.

The overview of this paper is as follows.
The section~2 reviews the Einstein-Cartan pseudoscalaron with matter fields and introduces the $(\epsilon R)^2$ model \cite{Salvio:2022suk} effective for inflation.
In Sec.~3, we introduce the extension of the kinetic term of the fermion.
From this extension in Einstein-Cartan pseudoscalaron theory, the equation of motion(EoM) of fermions is non-trivial and thus non-thermal particle production occurs even in the FRLW universe.
In Sec.~4, several results of numerical calculations about non-thermal particle production are exhibited.
By these results, it can be concluded that more larger value of $\beta$ contributes to more particle production.
Also, it is observed that the behavior of the number density is very different depending on whether the mass of the fermion is lighter, heavier or intermediate compared to the inflaton mass.
In Sec.~5, we apply the particle production to reheat the universe.
Finally, a discussion and summary of this paper are presented in Sec.~6.

Calculations in this paper are based on the following notations.
$m_p = \sqrt[]{1/8\pi G_N}$ is the planck mass and $m_{\phi}$ is the inflaton mass.
$G_N$ means the gravitational Newton constant.
The gamma matrix is defiend by $\lbrace\gamma_i,\gamma_j\rbrace =2\eta_{ij}$ where $\eta_{ij}$ is the minkowski metric $\eta=\text{diag}(-,+,+,+)$.
The definition of gamma matrice is
\begin{equation}
  \gamma^{\mu} = -i \notag
  \begin{pmatrix}
    0 & \sigma^{\mu} \\
    \overline{\sigma}^{\mu} & 0 \notag
  \end{pmatrix},
\end{equation}
\begin{gather}
  \sigma^{\mu} = (1, \bm{\sigma}), \notag \\
  \overline{\sigma}^{\mu} = (1, -\bm{\sigma}), \notag
\end{gather} 
\begin{equation}
  \gamma^5 = i\gamma^0\gamma^1\gamma^2\gamma^3. \notag
\end{equation}
Charge conjugate matrix $\mathcal{C}$ is
\begin{equation}
  \mathcal{C} = i\gamma^2\gamma^0. \notag
\end{equation}
$\epsilon_{\mu\nu\rho\sigma}$ is Levi-Civita antisymmetric symbol $\epsilon^{0123}\equiv 1$.
We use the natural units($\hbar = c= 1$). 
The symmetrization $A_{\lbrace i,j\rbrace}$ and the anti-symmetrization $A_{[i,j]}$ are respectively $\frac{1}{2}(A_{ij} + A_{ji})$ and $\frac{1}{2}(A_{ij} - A_{ji})$.
Greek indices mean the coordinate of spacetime while Roman indices mean that of local Minkowski spacetime.
$\hat{\mathcal{O}}$ means a q-number.

\section{Einstein-Cartan pseudoscalaron inflation from modified Holst term}

EC gravity is a metric-compatible metric-affine theory \cite{Hehl:1994ue,Gronwald:1997bx}, where the metric and the connection are independent variables. Fundamental conditions in this theory are the tetrad hypothesis and the metric compatible condition,
\begin{align}
  \partial_{\mu}e^i_{\nu} + {\omega^i}_{j\mu}e^j_{\nu} - {\Gamma^{\alpha}}_{\nu\mu}e_{\alpha}^i = 0, \label{eq:metric-compatible} \\
  \nabla_{\mu}g_{\nu\rho} = 0,
\end{align}
where, $e^i_{\mu}$ represents the tetrad which satisfies $g_{\mu\nu}=e^i_{\mu}e^j_{\nu}\eta_{ij}$.
The inverse of $e^i_{\mu}$ serves as a basis component of local Minkowski spacetime, $e^{\mu}_ie^{\nu}_jg_{\mu\nu}=\eta_{ij}$.
${\Gamma^{\mu}}_{\nu\rho}$ is the affine connection and 
${\omega^{ij}}_{\mu}$ is the gauge field of local Lorentz transformation that is called spin connection.
Thus, the covariant derivative $\nabla$ of spacetime is defined by  $\nabla_{\mu}A^{\nu}\equiv \partial_{\mu}A^{\nu}+{\Gamma^{\nu}}_{\alpha\mu}A^{\alpha}$,
and the covariant derivative $D$ of local Minkowski spacetime is defined by $D_{\mu}B_i = \partial_{\mu}B_i + {{\omega_{i}}^j}_{\mu}B_j$.
By Eq.~\eqref{eq:metric-compatible}, the 
curvature tensor ${R^{\mu}}_{\nu\rho\sigma}$ and the strength of local Lorentz transformation ${R^{ij}}_{\mu\nu}$ are connected by
\begin{align}
  {R^{\mu}}_{\nu\rho\sigma} &= {\partial_{\rho}\Gamma^{\mu}}_{\nu\sigma} - \partial_{\sigma}{\Gamma^{\mu}}_{\nu\rho} + {\Gamma^{\mu}}_{\alpha\rho}{\Gamma^{\alpha}}_{\nu\sigma} - {\Gamma^{\mu}}_{\alpha\sigma}{\Gamma^{\alpha}}_{\nu\rho} \notag \\
                            &= e^{\mu}_ie_{\nu j}(\partial_{\rho}{\omega^{ij}}_{\sigma} - \partial_{\sigma}{\omega^{ij}}_{\rho} + {\omega^i}_{k\rho}{\omega^{kj}}_{\sigma} - {\omega^i}_{k\sigma}{\omega^{kj}}_{\rho}) \notag \\
                            &= e^{\mu}_ie_{\nu j}{R^{ij}}_{\rho\sigma}.
\end{align}
In this theory, the existence of the antisymmetric part of the affine connection(torsion) is not prohibited.
Thus, the connections $\Gamma$, $\omega$ and the curvature scalar $R$ can be generally separated into torsionless and torsionful parts,
\begin{align}
  {\Gamma^{\mu}}_{\nu\rho} = {\Gamma^{\diamond\mu}}_{\nu\rho} + {K^{\mu}}_{\nu\rho}, \\
  {\omega^{ij}}_{\mu} = {\omega^{\diamond ij}}_{\mu} + {K^{ij}}_{\mu}, \\
  R =R^{\diamond}+T-2\nabla^{\diamond}_{\mu}T^\mu,
\end{align}
with 
\begin{align}
    {T^\rho}_{\mu\nu}\equiv{\Gamma^\rho}_{\mu\nu}-{\Gamma^\rho}_{\nu\mu},
\end{align}
and
\begin{align}   
    T 
    = \frac{1}{4} T^{\rho\mu\nu} T_{\rho\mu\nu}
    - \frac{1}{2} T^{\rho\mu\nu} T_{\mu\nu\rho}
    - T^\mu T_\mu,
\end{align}
where the script $\diamond$ indicates the torsionless parts which consist of metric and tetrad, 
${\Gamma^{\diamond\mu}}_{\nu\rho}$ is Levi-Civita symbol and ${\omega^{\diamond ij}}_{\mu} \equiv -e^{j\nu}\nabla^{\diamond}_{\mu}e^i_{\nu}$.
The torsionful parts consist of the contorsion,  ${K^{\mu}}_{\nu\rho}$, defined by ${K^{\mu}}_{\nu\rho} \equiv \frac{1}{2}({T^{\mu}}_{\nu\rho} + {T_{\nu\rho}}^{\mu} + {T_{\rho\nu}}^{\mu})$ and ${K^{ij}}_{\mu} \equiv e^i_{\rho}e^{j\nu}{K^{\rho}}_{\nu\mu}$.

\begin{figure}[t]
  \centering
  \begin{minipage}[h]{0.9\linewidth}
    \centering
    \includegraphics[width=1\linewidth]{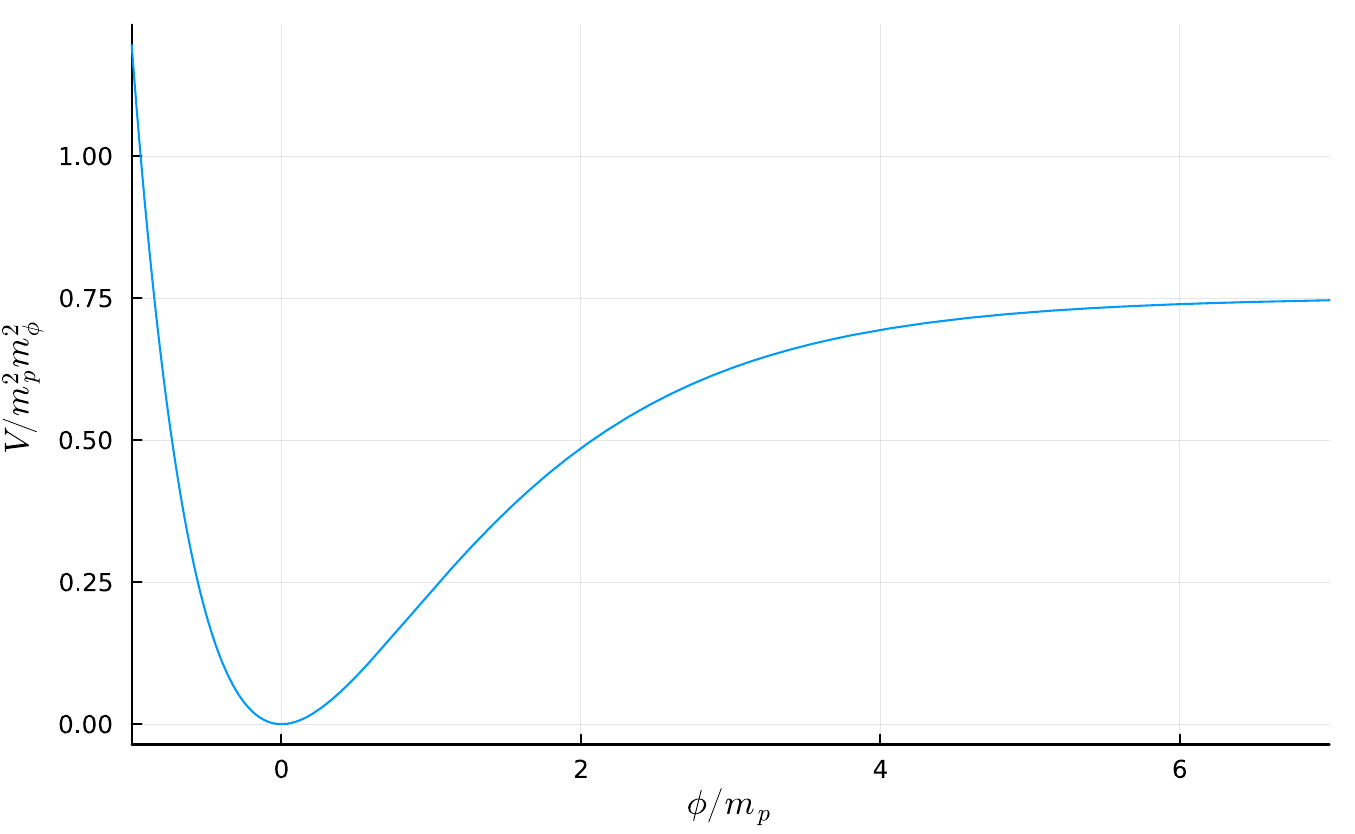}
  \end{minipage}
  \caption{The potential of $\phi$ in the $(\epsilon R)^2$ model with parameters $b=-320, m_{\phi}=\sqrt[]{\frac{d^2V}{d\phi^2}}|_{\phi=0}=\sqrt{\frac{(1+b^2)}{12c}}\sim2.76*10^{13} \text{GeV}$}
  \label{ref:potential}
\end{figure}

In our research, we start from the action with a function of the Holst term, $\epsilon R \equiv \epsilon^{\mu\nu\rho\sigma}R_{\mu\nu\rho\sigma}/e$,
\begin{align}
  S = \int ed^4x \Big\{
    \frac{m^2_p}{2}R + \frac{m^2_p}{4}H(\epsilon R) + \mathcal{L}_{\text{matter}}
    \Big\} , \label{eq:original_action}
\end{align}
where $e=\text{det}(e^i_{\mu})$.
Since the modification of the curvature scalar $R$ can not be represented by a dynamical scalar field \cite{Pradisi:2022nmh,Salvio:2022suk,He:2024wqv}, $F(R)$ regime is not considered.
Introducing an auxiliary field $\chi$, we obtain the action
\begin{align}
  S = \int ed^4x \Big\{
    \frac{m^2_p}{2}R + \frac{m^2_p}{4}H'(\chi)\epsilon R - V(\chi) + \mathcal{L}_{\text{matter}}
    \Big\} , \label{eq:original_action_aux}
\end{align}
where $V(\chi) = \frac{m^2_p}{4}(H'(\chi)\chi - H(\chi))$ and $H'(\chi) = \frac{dH(\chi)}{d\chi}$.
$\chi$ is a pseudoscalar field.
The action \eqref{eq:original_action} is reproduced by substituting the Euler-Lagrange equation with respect to $\chi$ into Eq.~\eqref{eq:original_action_aux}.
The Cartan equation is obtained as the Euler-Lagrange equation with respect to ${\omega^{ij}}_{\mu}$,
\begin{eqnarray}
  {T^{\mu}}_{ij} - T_ie^{\mu}_j + T_je^{\mu}_i + H'(\chi)\epsilon^{\mu\alpha\beta\gamma}e_{\delta[i}e_{j]\alpha}{T^\delta}_{\beta\gamma}/e \nonumber \\
  = {S^{\mu}}_{ij} + \epsilon^{\mu\alpha\beta\gamma}e_{\alpha i}e_{\beta j}(\partial_{\gamma}H'(\chi))/e, \label{eq:Cartanequation}
\end{eqnarray}
where $T_{\mu}\equiv {T^{\nu}}_{\mu\nu}$ is the torsion vector and ${S^{\mu}}_{ij}$~$\equiv$~$\frac{2}{m^2_p}\frac{\partial{\mathcal{L}}_{\text{matter}}}{\partial {{\omega}^{ij}}_{\mu}}$ is the spin density.
Thus, the torsion is rewritten in terms of tetrad $e^i_{\mu}$, matter field and auxiliary field $\chi$.
The effective metric theory is obtained by inserting the solution Eq.~\eqref{eq:Cartanequation} into Eq.~\eqref{eq:original_action_aux}.
For the vacuum (${S^{\mu}}_{mn}=0$), it becomes
\begin{equation}
  S = \int ed^4x\Big\{\frac{m^2_p}{2}R^{\diamond} - \frac{3m^2_p}{4}\frac{\partial_{\mu}H'\partial^{\mu}H'}{(1+H'^2)}-V(\chi)\Big\}.
\end{equation}
We introduce the pseudoscalaron $\phi$ by the redefinition, $H'(\chi) = \text{sinh}\Bigl(\frac{\sqrt[]{2}(\phi + \delta)}{\sqrt[]{3}m_p}\Bigr)$, and obtain
\begin{equation}
  S = \int ed^4x \Big\{\frac{m^2_p}{2}R^{\diamond} - \frac{1}{2}\partial_{\mu}\phi\partial^{\mu}\phi - V(\chi(\phi))\Big\},
\end{equation}
where $\delta$ is a constant to impose  $V(\phi=0)=0$.
For example, in the $(\epsilon R)^2$ model \cite{Salvio:2022suk}
\begin{equation}
  H(\epsilon R) = b \epsilon R + c(\epsilon R)^2,
\end{equation}
we obtain the pseudoscalaron with a potential, $V(\phi) = \frac{m^2_p}{16c}\Bigl(\sinh\Bigl(\frac{\sqrt[]{2}(\phi + \delta)}{\sqrt[]{3}m_p}\Bigr)-b\Bigr)^2$.
The constant $\delta$ is fixed to satisfy $\text{sinh}\Bigl(\frac{\sqrt[]{2}\delta}{\sqrt[]{3}m_p}\Bigr)=b$.
As is shown in Fig.\ref{ref:potential}, a certain value of parameters $b,c$ can realize the potential with a plateau.

\subsection{Pseudoscalaron inflation in FLRW universe}
We consider the homogeneous and isotropic universe described by the FLRW metric,
\begin{align}
  ds^2 = -dt^2 + a^2(t)(dx^2+dy^2+dz^2). \label{eq:FLRW1}
\end{align}
where $x^{\mu}=(t,x,y,z)$ represents the cosmological time and the conformal space, 
and $a(t)$ is the scale factor.
By introducing the conformal time, $d\eta=a^{-1}dt$, the spacetime is represented by
\begin{align}
  ds^2 = a^2(\eta)(-d\eta^2+dx^2+dy^2+dz^2), \label{eq:FLRW2}
\end{align}
The dot and the prime represent the derivative with respect to the cosmological time $t$, $\dot{()}=\frac{d}{dt}()$, and the conformal time $\eta$, $()'=\frac{d}{d\eta}()$, respectively.
The metric \eqref{eq:FLRW2} is used for the analysis in Sec.~4. 
In the metric \eqref{eq:FLRW1},
the scale factor is developed through the Friedmann-Robertson equations,
\begin{align}
  H^2 = \frac{\rho}{3m^2_p}, \label{eq:FR1}\\
  \dot{H} = -\frac{1}{2m^2_p}(\rho + p), \label{eq:FR2}
\end{align}  
where $H$ is the Hubble parameter defiend by $H=\frac{\dot{a}}{a}$, and $\rho$ and $p$ respectively denote the energy density and the pressure of the matter.
From Eqs.~\eqref{eq:FR1} and \eqref{eq:FR2}, we derive
\begin{align}
  \frac{\ddot{a}}{a} = -\frac{1}{6m^2_p}(\rho + 3p).
\end{align}
When the scalar field $\varphi$ distributes homogeneously,
the energy density and the pressure are described as $\rho=\frac{1}{2}\dot{\varphi}^2+V(\varphi)$ and $p=\frac{1}{2}\dot{\varphi}^2-V(\varphi)$.
The accelerated expansion takes place for $V(\varphi)>\dot{\varphi}^2$.

If the potential $V(\varphi)$ has a plateau, the scalr field starting from the plateau induces an inflationary expansion.
To solve the horizon and flatness problems encountered in the expanding universe,
the total e-folding number $N_e$, defined as $N_e=\log(a_f/a_i)$, should exceed $50\sim60$.
Here, $a_f$ represents the value of the scale factor at the end of inflation, while $a_i$ represents the value of the scale factor at the start of inflation.
To obtain the number, we often employ the slow-roll inflation scenario.
In the slow-roll approximation, the end of inflation is fixed by the slow-roll parameters $\varepsilon$ and $\eta$, defined by $\varepsilon=\frac{m^2_p}{2}\Bigl(\frac{1}{V}\frac{dV}{d\varphi}\Bigr)^2,\eta=m^2_p\frac{1}{V}\frac{d^2V}{d\varphi^2}$.

The quantum fluctuations of $\varphi$ induce the curvature perturbation, $P$, during inflation.
In terms of the conformal momentum space, a component of the fourieor decomposition of the curvature perturbation is represented as $P(k)=P_rk^{n_s}$.
Applying the slow-roll approximation, 
the amplitude $P_r$ of the curvature perturbation can be derived by $P_r=\frac{V(\varphi)}{24\pi^2m^4_p\varepsilon(\varphi)}$.
The spectral index $n_s$ is derived by $n_s = 1-6\varepsilon(\varphi) + 2\eta(\varphi)$.
The scalar-tensor ratio $r$ is calculated by $r=16\varepsilon(\varphi)$. 

The potential has a plateau in the $(\epsilon R)^2$ model with $b=-320, m_{\phi}=\sqrt[]{\frac{d^2V}{d\phi^2}}|_{\phi=0}=\sqrt{\frac{(1+b^2)}{12c}}\sim2.76*10^{13} \text{GeV}$ (Fig.\ref{ref:potential}),
and the slow-roll inflation scenario can be adopted .
We assume that the pseudoscalar field regarded as an inflaton dominates the energy density of the early universe.
Consequently, we obtain values such as $n_s\sim0.969, r\sim0.003, P_r\sim2.1*10^{-9}, N_e\sim60$ that agree with the observation regarding the Cosmic Microwave Background (CMB).
Below, we adapt this model to the non-thermal particle production after the end of inflation.
In our analysis, we consider that the particle production starts at $\phi\sim0.94m_p, \frac{d\phi}{dt}\sim-0.293m_pm_{\phi}$ where the slow-roll parameter $\varepsilon(\phi)$ becomes unity.

\subsection{The dynamics of background field after the end of inflation}
After the end of inflation, the oscillating inflaton $\phi$ dominates the energy density of the universe.
The potential is approximated to be $V\sim\frac{m^2_{\phi}}{2}\phi^2$ during the particle production.
The energy density of the inflaton and the scale factor are fixed by the Friedmann equations \eqref{eq:FR1} and \eqref{eq:FR2}.
Since the contribution to the pressure is cancelled between the kinetic and the potential energy,
$\rho_{\phi}$ follows the $\dot{\rho_{\phi}} + 3H\rho_{\phi}=0$.
The solution of this equation with \eqref{eq:FR1} is given by
\begin{align}
  \rho_{\phi} = m^2_pm^2_{\phi}\Bigl (\frac{\sqrt[]{3}}{2}m_{\phi}t + A\Bigr )^{-2}.\label{eq:matterden}
\end{align}
From the Eqs.~\eqref{eq:FR1} and \eqref{eq:matterden}, the scale factor is derived as
\begin{align}
   a(t) = A^{-\frac{2}{3}}\Bigl (\frac{\sqrt[]{3}}{2}m_{\phi}t + A \Bigr )^{\frac{2}{3}} \label{eq:scale_factor}.
\end{align}
The relation between cosmic time and conformal time is determined by $dt=ad\eta$ and $a(t=0)=a(\eta=0)=1$,
\begin{equation}
  \Bigl (\frac{\sqrt[]{3}}{2}m_{\phi}\eta + 3A \Bigr ) = 3A^{\frac{2}{3}}\Bigl (\frac{\sqrt[]{3}}{2}m_{\phi}t + A\Bigr )^{\frac{1}{3}}. \label{eq:concos}
\end{equation}
We assume that the oscillating part of the inflaton can be factored out,
\begin{align}
  \phi(t)/m_p &= CH(t)/m_{\phi}\sin(m_{\phi}t + B) \notag\\
              &=\frac{C}{ \sqrt[]{3}\Bigl (\frac{\sqrt[]{3}}{2}m_{\phi}t + A\Bigr )}\sin(m_{\phi}t + B). \label{eq:dynamicsOfphi}
\end{align}
It should be noted that Eq.~\eqref{eq:dynamicsOfphi} satisfies the EoM of the inflaton $\ddot{\phi} + 3H\dot{\phi} + \frac{dV}{d\phi} \simeq \ddot{\phi} + 3H\dot{\phi} + m^2_{\phi}\phi = 0$.
Arbitrary constants $A$, $B$ and $C$ are fixed by the initial values of $\phi$ and $\dot{\phi}$,
\begin{align}
  \rho_{\phi}|_{t=0}/m^2_pm^2_{\phi} &= A^{-2} \notag \\
                        &=\Bigl (\frac{1}{2}{\dot{\phi}^2} + \frac{1}{2}m^2_{\phi}{\phi^2}\Bigr )|_{t=0}/m^2_pm^2_{\phi},
\end{align}
\begin{equation}
  \cot(B) = \dot{\phi}/\phi|_{t=0} + \frac{\sqrt[]{3}}{2A},
\end{equation}
\begin{equation}
  \phi|_{t=0} = \frac{C}{ \sqrt[]{3}A}\sin B.
\end{equation}
We employ these formula as a simple background for the universe after the end of inflation.

\section{A model of the non-minimal couplings to fermion}
In Einstein-Cartan gravity, the matter action is generalized with non-minimal gravitational interactions. We consider a general fermion  Lagrangian constructed with operators up to four dimensions \cite{Freidel:2005sn},
\begin{widetext}
\begin{align}
  \mathcal{L}_{\text{matter}}
                       =-\frac{1}{2}(\overline{\psi}(1 - i\alpha - i\beta \gamma^5)\gamma^{\mu}(\partial_{\mu} + \frac{1}{4}{\omega^{ij}}_{\mu}\gamma_{ij})\psi+\text{h.c.})  - m_{\psi}\overline{\psi}\psi, \label{eq:non_minimal}
\end{align}
\end{widetext}
where the Dirac conjugate $\overline{\psi}$ is defined by $\overline{\psi}=i\gamma^{0}\psi^{\dagger}$ and h.c. means the hermitian conjugate.
This Lagrangian reduces to the ordinary one in the absence of gravity.
The extension causes the parity violation
 observed in the various astrophysical and elementary particle phenomena.
 
The spin density is given by
\begin{align}
  {S^{\mu}}_{ij}&=\frac{2}{m^2_p}\frac{\partial \mathcal{L}_{\text{matter}}}{\partial {\omega^{ij}}_{\mu}} \notag \\
                &=-\frac{1}{m^2_p}(\epsilon_{ijkl}A^le^{\mu k} - 2e^{\mu}_{[i}(\alpha V_{j]}+\beta A_{j]})).
\end{align}
$A_i \equiv \bar\psi i \gamma^5 \gamma_i \psi$ and $V_i \equiv \bar\psi i \gamma_i \psi$ denote the axial vector and the vector current, respectivly.
Performing the partial integration, the Lagrangian density \eqref{eq:non_minimal} can be decomposed into
\begin{equation}
  \mathcal{L}_{\text{matter}} = \mathcal{L}^{\diamond}_{\text{matter}} + \frac{1}{8}\hat{T}^{\mu}A_\mu - \frac{1}{2}T^{\mu}(\alpha V_{\mu} + \beta A_{\mu}),
\end{equation}
where the torsionless part is
\begin{eqnarray}
  \mathcal{L}^{\diamond}_{\text{matter}} = &-&\overline{\psi}\gamma^{\mu}\partial_{\mu}\psi- \frac{1}{2}\overline{\psi}\omega^{\diamond ijk}\eta_{ik}\gamma_j\psi
  \notag \\
  &-& \frac{i}{4}\epsilon_{ijkl}\omega^{\diamond ijk}\overline{\psi}\gamma^5\gamma^l\psi -m_{\psi}\overline{\psi}\psi. \label{FreeFermiPart}
\end{eqnarray}
Since non-minimal coupling parameters don't appear in the torsionless part~\eqref{FreeFermiPart}, these parameters only contribute to the interaction between the torsion and the fermion. 

By solving Eq.~\eqref{eq:Cartanequation}, the torsion is represented by the inflaton and the fermion,
\begin{align}
  T_{ijk} = -\frac{2}{3}\eta_{i[j}\eta_{k]l}T^l + \frac{1}{6}\epsilon_{ijkl}\hat{T}^l, \label{eq:solution}
\end{align}
with
\begin{eqnarray}
    T^l = \frac{\text{sech}^2(X(\phi))}{2}&\Bigl(&\frac{3}{2m^2_p}(\alpha V^l + (\beta + \sinh{X(\phi)})A^l) \notag \\ &+& 3\sinh{X(\phi)}\partial^l(\sinh{X(\phi)})\Bigr) ,\label{eq:torvec}
\end{eqnarray}
\begin{eqnarray}
  \hat{T}^l = -3~\text{sech}^2(X(\phi))\Bigl(\frac{1}{m^2_p}((-1+\sinh{X(\phi)}\beta)A^l \nonumber\\ + \sinh{X(\phi)}\alpha V^l) - 2\partial^l(\sinh{X(\phi)}) \Bigr),\label{eq:toraxi}
\end{eqnarray}
where we write $X(\phi) = \frac{\sqrt[]{2}(\phi+\delta)}{\sqrt[]{3}m_p}$.
Inserting the solution \eqref{eq:solution} into \eqref{eq:original_action_aux}, we obtain the effective metric action,
\begin{eqnarray}
  S = 
  \int e d^4x &\Big\{&
    \frac{m^2_p}{2}R^{\diamond} 
    - \frac{1}{2}\partial_{\mu}\phi\partial^{\mu}\phi - V(\phi) + \mathcal{L}^{\diamond}_{\psi}\nonumber\\
    &+& f_{\mu}(\phi)A^{\mu} + g_{\mu}(\phi)V^{\mu}
    +\mathcal{L}_{\text{4-fermi}}
    \Big\}, \label{eq:eff_met_action}
\end{eqnarray}
with
\begin{align}
  f_{\mu} = \frac{\sqrt[]{3}}{2\ \sqrt[]{2}}\partial_{\mu}\Bigl(\frac{\phi}{m_p}\Bigr) 
          \Bigl\{ -\text{sech}(X(\phi)) + \beta \tanh(X(\phi))\Bigr\}, \label{eq:inf_dyn1}\\
  g_{\mu} = \frac{\sqrt[]{3}\alpha}{2\ \sqrt[]{2}}\partial_{\mu}\Bigl(\frac{\phi}{m_p}\Bigr)\tanh(X(\phi)) ,\label{eq:inf_dyn2}
\end{align}
and four-fermion interactions,
\begin{widetext}
    \begin{align}
  \mathcal{L}_{\text{4-fermi}}
  = \frac{3~\text{sech}^2(X(\phi))}{16m^2_p}
                                \Bigl(\alpha^2V^2 + 2(\alpha\beta + \alpha~\text{sinh}(X(\phi)))VA + (\beta^2 +2\beta\text{sinh}(X(\phi)) - 1)A^2 \Bigr),\label{eq:four-fermi}
\end{align}
\end{widetext}
where $A^2$, $VA$ and $V^2$ are defined by $A^2=A^{\mu}A_{\mu}$, $VA=V^{\mu}A_{\mu}$ and $V^2=V^{\mu}V_{\mu}$.

The four-fermion interactions, \eqref{eq:four-fermi}, are suppressed by the factor $m^{-2}_p$.
On the other hand, the five-dimensional interactions,
\begin{align}
    \mathcal{L}_{\phi\psi\psi} = f_{\mu}(\phi)A^{\mu} + g_{\mu}(\phi)V^{\mu},
\end{align}
are suppressed by $m^{-1}_p$.
Therefore, $\mathcal{L}_{\phi\psi\psi}$ becomes the leading order term in the reheatig era.
We neglect the higher order terms and apply the formalism developed in the previous works of the fermionic preheating \cite{Greene:1998nh,Greene:2000ew,Giudice:1999fb,Peloso:2000hy,Herring:2020cah}.

In the previous works, Yukawa type interaction, 
\begin{align}
    \phi\overline{\psi}\psi,
\end{align}
is considered.
In our model, the form of the interactions, 
\begin{align}
    \partial_{\mu}\phi\overline{\psi}i\gamma^{\mu}\psi, \partial_{\mu}\phi\overline{\psi}i\gamma^5\gamma^{\mu}\psi,
\end{align}
are different from the Yukawa type interaction and
 violate the parity.

\subsection{The EoM in FLRW universe}
For discussing the fermionic non-thermal particle production after the end of inflation, we derive the EoM of the classical fermionic field $\psi$ and the Heisenberg operator $\hat{\psi}$.
In the spatially flat and homogeneous FLRW metric \eqref{eq:FLRW2}, the tetrad is given by $e^i_{\mu} = a\delta^i_{\mu}$ from its definition $g_{\mu\nu}=\eta_{ij}e^i_{\mu}e^j_{\nu}$, 
and components regarding the torsionless spin connection are $\eta_{ik}\omega^{\diamond ijk}= 3\frac{\mathcal{H}}{a}\delta^{j0}$ and $\epsilon_{ijkl}\omega^{\diamond ijk}=0$.
Thus, Eq.~\eqref{FreeFermiPart} becomes
\begin{align}
  \mathcal{L}^{\diamond}_{\text{matter}} = -\overline{\psi}\gamma^{\mu}\partial_{\mu}\psi - \frac{3\mathcal{H}}{2a}\overline{\psi}\gamma_0\psi
                                 -m_{\psi}\overline{\psi}\psi,
\end{align}
where $\mathcal{H}$ is defined by $\mathcal{H} = \frac{a^{\prime}}{a}$.
Therefore, the Lagrangian density of $\psi$ is denoted as
\begin{eqnarray}
  e\mathcal{L}_{\text{matter}} = 
    a^4 \Big\{
    &-&\frac{1}{a}\overline{\psi}\delta^{\mu}_i\gamma^{i}\partial_{\mu}\psi 
    - \frac{3\mathcal{H}}{2a}\overline{\psi}\gamma_0\psi
    - m_{\psi} \overline{\psi}\psi 
    \nonumber \\ &+& \frac{1}{a}\delta^{\mu}_i (f_{\mu}A^{i} + g_{\mu}V^i)
  \Big\},
  \label{eq:fermionic_action}
\end{eqnarray}
where $f_{\mu}(\phi)$ and $g_{\mu}(\phi)$ are functions of the inflaton defined by Eqs.~\eqref{eq:inf_dyn1} and \eqref{eq:inf_dyn2}.
Rescaling the fermion as $a^{\frac{3}{2}}\psi \rightarrow \psi$, one can finally obtain the Lagrangian density,
\begin{align}
  e\mathcal{L}_{\text{matter}} = 
    -\overline{\psi}\delta^{\mu}_i\gamma^{i}\partial_{\mu}\psi 
    -m_{\psi}a \overline{\psi}\psi + \delta^{\mu}_i (f_{\mu}A^{i} + g_{\mu}V^i). \label{eq:fermionic_Lagrangian}
\end{align}
The inflaton $\phi$ after the end of inflation is assumed to be a homogeneous field, $\partial_{x}\phi = \partial_{y}\phi = \partial_{z}\phi = 0$.
The EoM of $\psi$ is
\begin{align}
  -\gamma^{\mu}\partial_{\mu}\psi
  -m_{\psi}a \psi + i(g_{0}\gamma^0 + f_{0}\gamma^5\gamma^0)\psi = 0. \label{eq:EoMofpsi}
\end{align}
We note that $\gamma^{\mu}$ in Eq.~\eqref{eq:EoMofpsi} is the gamma matrices in the local Lorentz frame.
Since the Heisenberg operator $\hat{\psi}$ also satisfies the identical equation,
\begin{align}
  -\gamma^{\mu}\partial_{\mu}\hat{\psi}
  -m_{\psi}a \hat{\psi} + i(g_{0}\gamma^0 + f_{0}\gamma^5\gamma^0)\hat{\psi} = 0,\label{eq:heisenberg}
\end{align}
we can obtain the EoM of the spinors $u_s(\bm{k},\eta)$ and $v_s(\bm{k},\eta)$ by the decomposition of the operator $\hat{\psi}$,
\begin{align}
  &\hat{\psi} = \int \frac{d^3\bm{k}}{\sqrt[]{2}(2\pi)^3} 
  \nonumber \\ &\sum_{s}\lbrack u_s(\bm{k},\eta)\hat{a}_s(\bm{k})e^{i\bm{k}\bm{x}} + v_s(\bm{k},\eta)\hat{b}^{\dagger}_s(\bm{k})e^{-i\bm{k}\bm{x}}\rbrack, \label{eq:decomposition}
\end{align}
where $\hat{a}_s(\bm{k}),\hat{b}_s(\bm{k})$ indicate (anti) particle annihilation operator.
They satisfy the anti-commutation relations,
\begin{gather}
  \lbrace \hat{a}_s(\bm{k}), \hat{a}^{\dagger}_r(\bm{l}) \rbrace = (2\pi)^3\delta^3(\bm{k}-\bm{l})\delta_{rs}, \label{eq:anti_commutation1}\\
  \lbrace \hat{b}_s(\bm{k}), \hat{b}^{\dagger}_r(\bm{l}) \rbrace = (2\pi)^3\delta^3(\bm{k}-\bm{l})\delta_{rs}, \label{eq:anti_commutation2}\\
  \lbrace \hat{a}_s(\bm{k}), \hat{a}_r(\bm{l}) \rbrace = 0,\label{eq:anti_commutation3} \\
  \lbrace \hat{b}_s(\bm{k}), \hat{b}_r(\bm{l}) \rbrace = 0.
  \label{eq:anti_commutation4}
\end{gather}
We consider the case where the charge conjugate of fermion, 
\begin{align}
\mathcal{C}\overline{\hat{\psi}}^T=i\gamma^2\gamma^0\overline{\hat{\psi}}^T,
\end{align}
is well-defined and interchanges the particle and the antiparticle.
Then,
we obtain the relation between $u_s(\bm{k},\eta)$ and $v_s(\bm{k},\eta)$,
\begin{align}
  u_s(\bm{k},\eta) = \mathcal{C} \overline{v_s(\bm{k},\eta)}^T. \label{eq:spinor}
\end{align}
From Eq.~\eqref{eq:heisenberg} and its charge conjugate with \eqref{eq:decomposition} and \eqref{eq:spinor}, $v_s(-\bm{k},\eta)$ should satisfy
\begin{gather}
  (-\gamma^0\partial_0 - i\gamma\bm{k} -am_{\psi} + i(g_{0}\gamma^0 + f_{0}\gamma^5\gamma^0))v_s(-\bm{k},\eta) = 0, \\
  (-\gamma^0\partial_0 - i\gamma\bm{k} -am_{\psi} + i(-g_{0}\gamma^0 + f_{0}\gamma^5\gamma^0))v_s(-\bm{k},\eta) = 0.
\end{gather}
Since $g_0$ is proportional to $\alpha$, these equations are satisfied for $\alpha=0$. 

We define the spinor $u_s(\bm{k},\eta)$ as
\begin{equation}
  u_s(\bm{k},\eta) = 
  \begin{pmatrix}
    u^+_{s,k}(\eta)\xi_{s,\bm{k}} \\
    u^-_{s,k}(\eta)\xi_{s,\bm{k}}
  \end{pmatrix},\label{eq:assumption}
\end{equation}
where $s$ indicates the spin direction and $\xi_{s,\bm{k}}$ describes the eigen-spinor of helicity,
\begin{equation}
  \bm{\sigma}\bm{k}\xi_{s,\bm{k}} = s k\xi_{s,\bm{k}},
\end{equation}
\begin{equation}
  \xi_{s,-\bm{k}} = -i\sigma_2\xi^{\ast}_{s,\bm{k}}.
\end{equation}
The equation \eqref{eq:heisenberg} is rewritten as
\begin{align}
  \begin{aligned}
    u^{\pm}_{s,k}(\eta)' = \pm i(ks + f_0)u^{\pm}_{s,k}(\eta) - iam_{\psi}u^{\mp}_{s,k}(\eta).
  \end{aligned}
  \label{eq:initial_velocities}
\end{align}
Performing the time derivative, we obtain the EoMs of the amplitude of the spinor, $u^{+,-}_{s,k}(\eta)$,
\begin{eqnarray}
    u^{\pm}_{s,k}(\eta)'' = -((ks + f_0)^2 &+& a^2m^2_{\psi} \mp if_0')u^{\pm}_{s,k}(\eta) \nonumber \\ &-& ia'm_{\psi}u^{\mp}_{s,k}(\eta).
  \label{eq:equations_amplitude}
\end{eqnarray}
The EoM~\eqref{eq:heisenberg} guarantees the relation $|u^{+}_{s,k}|^2 + |u^{-}_{s,k}|^2 = 2$ with the anti-commutation relations~\eqref{eq:anti_commutation1}-\eqref{eq:anti_commutation4}, the assumption~\eqref{eq:assumption} and the canonical anti-commutation relation $\{\psi(\eta, \bm{x}), \psi^{\dagger}(\eta, \bm{y})\}=\delta^3(\bm{x}-\bm{y})$. 

Below, we evaluate the particle production for $\alpha=0$.

\subsection{The number density of fermion}
From the Lagrangian density \eqref{eq:fermionic_Lagrangian}, we can define the Hamiltonian operator $\hat{H}_{\psi}$,
\begin{align}
  \hat{H}_{\psi} = \frac{1}{a}\int d^3x(\overline{\hat{\psi}}\gamma^{I}\partial_{I}\hat{\psi}
  + m_{\psi}a \overline{\hat{\psi}}\hat{\psi} - f_{0}(\phi) \overline{\hat{\psi}}i\gamma^5\gamma^0\hat{\psi}), \label{eq:Hamiltonian}
\end{align}
where the script $I$ denotes the space components, $I\in\{x,y,z\}$, and the factor $a^{-1}$ comes from the fact that the Hamiltonian is the generator of the translation regarding the cosmic time $t$, and the relation $dt = a(\eta)d\eta$.
By inserting \eqref{eq:decomposition} into \eqref{eq:Hamiltonian}, 
the Hamiltonian becomes
\begin{widetext}
\begin{align}
  \hat{H}_{\psi} = \frac{1}{a}\int \frac{d^3\bm{k}}{(2\pi)^3} \sum_s 
  \lbrack E^s_k(\hat{a}^{\dagger}_s(\bm{k})\hat{a}_s(\bm{k}) - \hat{b}_s(\bm{k})\hat{b}^{\dagger}_s(\bm{k})) + F^s_k \hat{b}_s(-\bm{k})\hat{a}_s(\bm{k}) + F^{s\ast}_k \hat{b}^{\dagger}_s(-\bm{k})\hat{a}^{\dagger}_s(\bm{k}) \rbrack,
  \label{eq:hamiltomian_operator}
\end{align}
\end{widetext}
where the coefficients $E^s_k$ and $F^s_k$ are
\begin{gather}
  E^s_k = (ks + f_0)(1 - |u^{+}_{s,k}|^2) + am_{\psi} \text{Re}(u^{+}_{s,k}u^{-\ast}_{s,k}), \\
  F^s_k = -(ks + f_0)u^{+}_{s,k}u^{-}_{s,k} + am_{\psi}(- u^{+}_{s,k}u^{+}_{s,k} + u^{-}_{s,k}u^{-}_{s,k}).
\end{gather}
 Since they satisfy $|E^s_k|^2 + |F^s_k|^2 = \omega^{s2}_k = a^2m^2_{\psi} + (ks + f_0)^2$, 
the Hamiltonian operator \eqref{eq:hamiltomian_operator} is diagonalyzed by introducing the time-dependent annihilation operators, $\hat{a}_s(\bm{k}, \eta)$ and $ \hat{b}_s(\bm{k}, \eta)$,
 \begin{equation}
  \hat{H}_{\psi} = \int\frac{d^3\bm{k}}{(2\pi)^3} \sum_s 
  \frac{\omega^s_k}{a}(\hat{a}^{\dagger}_s(\bm{k}, \eta)\hat{a}_s(\bm{k}, \eta) - \hat{b}_s(\bm{k}, \eta)\hat{b}^{\dagger}_s(\bm{k}, \eta)). 
 \end{equation}
 The Bogoliubov transformation connects the operators $\hat{a}_s(\bm{k}),\hat{b}^{\dagger}_s(-\bm{k})$ and $\hat{a}_s(\bm{k}, \eta),\hat{b}^{\dagger}_s(\bm{k}, \eta)$,
\begin{align}
  \begin{pmatrix}
    \hat{a}_s(\bm{k}, \eta) \\ \hat{b}^{\dagger}_s(\bm{k}, \eta)
  \end{pmatrix}
  = 
  \begin{pmatrix}
    \alpha_s(k, \eta) & \beta_s(k, \eta) \\
    -\beta^{\ast}_s(k, \eta) & \alpha^{\ast}_s(k,\eta)
  \end{pmatrix}
  \begin{pmatrix}
    \hat{a}_s(\bm{k}) \\ \hat{b}^{\dagger}_s(-\bm{k})
  \end{pmatrix},
\end{align}
where $\alpha_s(k, \eta)$ and $\beta_s(k, \eta)$ are the Bogoliubov coefficients defined by
\begin{gather}
  |\alpha_s(k, \eta)|^2 = \frac{\omega^s_k + E^s_k}{2\omega^s_k}, \\
  |\beta_s(k, \eta)|^2 = \frac{\omega^s_k - E^s_k}{2\omega^s_k}.
\end{gather}
From the definition of the Bogoliubov coefficients and the anti-commutation relations \eqref{eq:anti_commutation1}-\eqref{eq:anti_commutation4},
the anti-commutation relations for the operators, $\hat{a}_s(\bm{k}, \eta), \hat{b}^{\dagger}_s(-\bm{k}, \eta)$, are derived to be
\begin{gather}
  \lbrace \hat{a}_s(\bm{k},\eta), \hat{a}^{\dagger}_r(\bm{l},\eta) \rbrace = (2\pi)^3\delta^3(\bm{k}-\bm{l})\delta_{rs}, \\
  \lbrace \hat{b}_s(\bm{k},\eta), \hat{b}^{\dagger}_r(\bm{l},\eta) \rbrace = (2\pi)^3\delta^3(\bm{k}-\bm{l})\delta_{rs}, \\
  \lbrace \hat{a}_s(\bm{k},\eta), \hat{a}_r(\bm{l},\eta) \rbrace = 0, \\
  \lbrace \hat{b}_s(\bm{k},\eta), \hat{b}_r(\bm{l},\eta) \rbrace = 0.
\end{gather}
We redefine the Hamiltonian operator so that a minimum of its expectation value is zero,
\begin{equation}
  \hat{H}_{\psi} = \int\frac{d^3\bm{k}}{(2\pi)^3} \sum_s
  \frac{\omega^s_k}{a}(\hat{a}^{\dagger}_s(\bm{k}, \eta)\hat{a}_s(\bm{k}, \eta) + \hat{b}^{\dagger}_s(\bm{k}, \eta)\hat{b}_s(\bm{k}, \eta)). \label{eq:new_hamiltonian}
 \end{equation}
 
We consider the vacuum state defined by $\hat{a}_s(\bm{k})(,\hat{b}_s(\bm{k}))|0\rangle_{\eta_0}=0$ at $\eta=\eta_0$ .
Under the state, the expectation value of the number operators are developed
\begin{align}
  \Bigl \langle 0 \Bigl |_{\eta_0} \hat{a}^{\dagger}_s(\bm{k}, \eta)\hat{a}_s(\bm{k}, \eta) \Bigr | 0 \Bigr \rangle_{\eta_0} 
                    &=\Bigl \langle 0 \Bigl |_{\eta_0} \hat{b}^{\dagger}_s(\bm{k}, \eta)\hat{b}_s(\bm{k}, \eta) \Bigr | 0 \Bigr \rangle_{\eta_0} \notag\\
                    &=|\beta_s(k, \eta)|^2. \label{eq:ExpVal}
\end{align}
The behavior of the expectation value of the number operator for the particle and the antiparticle is equivalent.
Even if the expectation value of Hamiltonian $\langle \hat{H}_{\psi}(\eta) \rangle_{\eta_0}$ is zero at $\eta_0$, it is not necessary to be so at $\eta > \eta_0$.
This means a non-thermal particle production.
From the conditions of the amplitude of the spinor,
\begin{gather}
  |u^{+}_{s,k}|^2 + |u^{-}_{s,k}|^2 = 2, \\
  \langle \hat{H}_{\psi}(\eta_0) \rangle_{\eta_0} = 0,
\end{gather}
the initial values of the amplitude of the spinor are derived,
\begin{gather}
  u^{+}_{s,k}(0) = \sqrt[]{1 - \frac{(ks + f_0(\phi(0)))}{\omega^s_k(0)}}, \label{eq:initial_conditio_1}\\
  u^{-}_{s,k}(0) = \frac{m_{\psi}a(0)}{\omega^s_k(0)}\Bigl( 1 - \frac{(ks + f_0(\phi(0)))}{\omega^s_k(0)} \Bigr)^{-\frac{1}{2}}. \label{eq:initial_conditio_2}
\end{gather}

We define the quantities to observe whether the particle production occurs.
The expectation value of total number density is given by
\begin{widetext}
\begin{align}
 \Bigl \langle 0 \Bigl |_{\eta_0} \frac{\hat{N}^s_{particle}}{a^3V} \Bigr | 0 \Bigr \rangle_{\eta_0} & \equiv \int \frac{d^3\bm{k}}{(2\pi)^3a^3V} \Bigl \langle 0 \Bigl |_{\eta_0} \hat{a}^{\dagger}_s(\bm{k}, \eta)\hat{a}_s(\bm{k}, \eta) \Bigr | 0 \Bigr \rangle_{\eta_0} = \frac{1}{2\pi^2a^3}\int dk k^2 \Bigl( \frac{1}{2} - \frac{E^s_k}{2\omega^s_k} \Bigr), \label{eq:NumberDensity}
\end{align}
\end{widetext}
where $V$ is the volume of conformal space.
The number density regarding the conformal momentum space is
\begin{equation}
  n^s_{m_{\psi},k} \equiv \frac{1}{2} - \frac{E^s_k}{2\omega^s_k} . \label{eq:NumberDensity_momentum}
\end{equation}
It vanishes at $\eta=\eta_0$ and must not exceed unity at anytime by Pauli blocking.
The total number of (anti) particles is defined by
\begin{align}
  N^s_{\psi} \equiv \int dk\frac{k^2 n^s_{m_{\psi},k}}{2\pi^2}.
\end{align}
We evaluate the energy density,
\begin{gather}
  \rho_{m_{\psi}} = 2 \Sigma_s\int dk \rho^s_{m_{\psi},k}, \label{eq:eneden_fermion} \\
  \rho^s_{m_{\psi},k} = \frac{k^2 \omega^s_k n^s_{m_{\psi},k}}{2\pi^2 a^4},
\end{gather} 
where the factor 2 comes from the degree of freedom of the particle and the antiparticle.

\section{Analytical and numerical results}
In this section, an analytical implication regarding the behavior of the number density and some numerical results are exhibited.
The former suggests the behavior for particles lighter than the inflaton. We will attempt a specific application of the latter results in Sec.~5.

\subsection{Massless limit}
We can analytically evaluate the particle production for a simple case.
Here, we consider the massless limit, $m_{\psi}=0$, which has several differences from the massive particle.

First, the initial conditions \eqref{eq:initial_conditio_1} and \eqref{eq:initial_conditio_2} are not appropriate for the massless limit.
The initial conditions of $u^{+}_{s,k}$ and $u^{-}_{s,k}$ are divided into four cases.
For $ f_0(0) > 0$, the initial conditions become
\begin{gather}
  u^{+}_{\uparrow, k}(0) = 0, \label{eq:massless1}\\
  u^{-}_{\uparrow, k}(0) = \sqrt[]{2}e^{i\theta},
\end{gather}
\begin{equation}
  u^{+}_{\downarrow, k}(0) =
  \begin{cases}
    \sqrt[]{2}e^{i\psi}, & \text{$k > |f_0(0)|$} \\
    0,                   & \text{$k < |f_0(0)|$}
  \end{cases}
\end{equation}
\begin{equation}
  u^{-}_{\downarrow, k}(0) =
  \begin{cases}
    0,                   & \text{$k > |f_0(0)|$} \\
    \sqrt[]{2}e^{i\psi}, & \text{$k < |f_0(0)|$}
  \end{cases}
\end{equation}
where $\theta$ and $\psi$ are arbitrary phases. For $f_0(0) < 0$, these are given by
\begin{gather}
  u^{+}_{\downarrow, k}(0) = \sqrt[]{2}e^{i\psi}, \\
  u^{-}_{\downarrow, k}(0) = 0,
\end{gather}
\begin{equation}
  u^{+}_{\uparrow, k}(0) =
  \begin{cases}
    0,                     & \text{$k > |f_0(0)|$} \\
    \sqrt[]{2}e^{i\theta}, & \text{$k < |f_0(0)|$}
  \end{cases}
\end{equation}
\begin{equation}
  u^{-}_{\uparrow, k}(0) =
  \begin{cases}
    \sqrt[]{2}e^{i\theta}, & \text{$k > |f_0(0)|$} \\
    0.                   & \text{$k < |f_0(0)|$}
  \end{cases}
  \label{eq:massless2}
\end{equation}
Second, the number densities are represented by
\begin{align}
    n_{\uparrow,k}(\eta) &= \frac{1}{2}(1 - \text{sign}(k + f_0)(1 - |u^+_{\uparrow, k}(\eta)|^2)), \label{eq:masslessnumden}
\end{align}
\begin{align}
    n_{\downarrow,k}(\eta) &= \frac{1}{2}(1 - \text{sign}(k - f_0)(1 - |u^-_{\downarrow, k}(\eta)|^2)). \label{eq:masslessnumden2}
\end{align}
Third, the 1st order differential equations \eqref{eq:initial_velocities} reduce to
\begin{gather}
  u^{\pm}_{s,k}(\eta)' = \pm i(ks + f_0(\eta))u^{\pm}_{s,k}(\eta). \label{eq:1stODE}
\end{gather}

From Eq.~\eqref{eq:1stODE}, the time derivative of $|u^{\pm}_{s,k}(\eta)|^2$ vanishes,
\begin{equation}
  |u^{\pm}_{s,k}(\eta)|^{2\prime} = 0.
\end{equation}
Because of a damped oscillation of the inflaton $\phi$, $|f_0|$ becomes smaller than $k$ after a sufficient amount of time.
The signature of $k+sf_0$ is positive and Eqs.~\eqref{eq:masslessnumden} and \eqref{eq:masslessnumden2} are simplified to
\begin{align}
  n_{\uparrow,k} &= \frac{1}{2}(1 - (1 - |u^+_{\uparrow, k}|^2)) \notag \\
                  &= \frac{1}{2}|u^+_{\uparrow, k}|^2_{\eta=0}, \label{eq:massless_numden1} \\
  n_{\downarrow,k} &= \frac{1}{2}(1 - (1 - |u^-_{\downarrow, k}|^2)) \notag \\
                  &= \frac{1}{2}|u^-_{\downarrow, k}|^2_{\eta=0}. \label{eq:massless_numden2}
\end{align}
Thus, the number densities are fixed by the initial values of $|u^{\pm}_{s, k}|^2_{\eta=0}$.
The non-thermal excitations strongly depend on the initial conditions at the massless limit.
The hericity of the produced particle is determined by the initial condition $f_0(0)$
(Table.\ref{table:masslessProduction}).
\begin{table}[t]
  \centering
   \begin{tabular}{|c|c|c|}
    \hline
                  & $k<|f_0(0)|$ & $k>|f_0(0)|$ \\ \hline
    $f_0(0)>0$ & $n^{\uparrow}_k=0$ and $n^{\downarrow}_k=1$ & $\times$ \\ \hline
    $f_0(0)<0$ & $n^{\uparrow}_k=1$ and $n^{\downarrow}_k=0$ & $\times$ \\ \hline
   \end{tabular}
  \caption{particle production in massless limit}
  \label{table:masslessProduction}
\end{table}
From the numerical simulation, it is observed that lighter particles have a similiar property.
If the reheating starts at $\phi^{\prime}(0)=0$, 
the production of lighter particles hardly occur due to the property in Tab.~\ref{table:masslessProduction} with $f_0(0) \propto \phi^{\prime}(0) = 0$.

\subsection{Numerical results}
Since it is difficult to solve Eq.~\eqref{eq:equations_amplitude} analytically,
we evaluate the behavior of the number density through numerical calculations.
Here, we set $\rho_{\psi}(0)=0$ and employ the $(\epsilon R)^2$ model for the dynamics of the inflaton governing the evolution of the universe.
As is shown in Sec.~2, initial values are fixed at  $\phi(0)\sim0.940m_p$ and $\frac{d\phi}{d\eta}(0)\sim-0.293m_pm_{\phi}$, and 
the inflaton dynamics is determined by the values $\phi(0)$, $\phi'(0)$.
We solve the Eq.~\eqref{eq:equations_amplitude} to obtain $n^s_{m_{\psi},k},N^s_{\psi}$ and $\rho_{m_{\psi}}$.
Numerical calculations in our research are performed by Julia and an algorithm specified by Verner9 solver in the package DifferentialEq.jl.
\subsubsection{Heavy and light fermion}
\begin{figure}[htbp]
  \centering
  \begin{minipage}[t]{0.9\linewidth}
      \centering
      \includegraphics[width=1\linewidth]{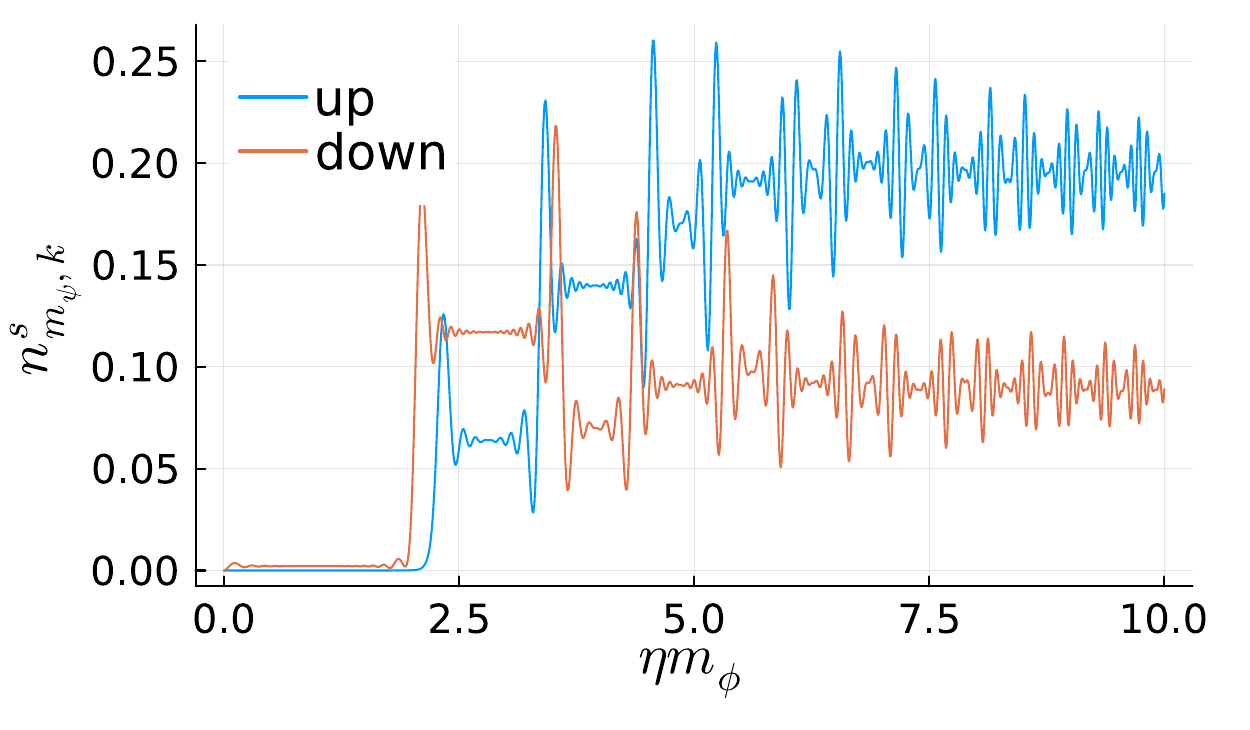}
      \subcaption{$m_{\psi}=4m_{\phi}$, $\bm{\beta=100}$, $k=10m_{\phi}$}
  \end{minipage}
  \begin{minipage}[t]{0.9\linewidth}
      \centering
      \includegraphics[width=1\linewidth]{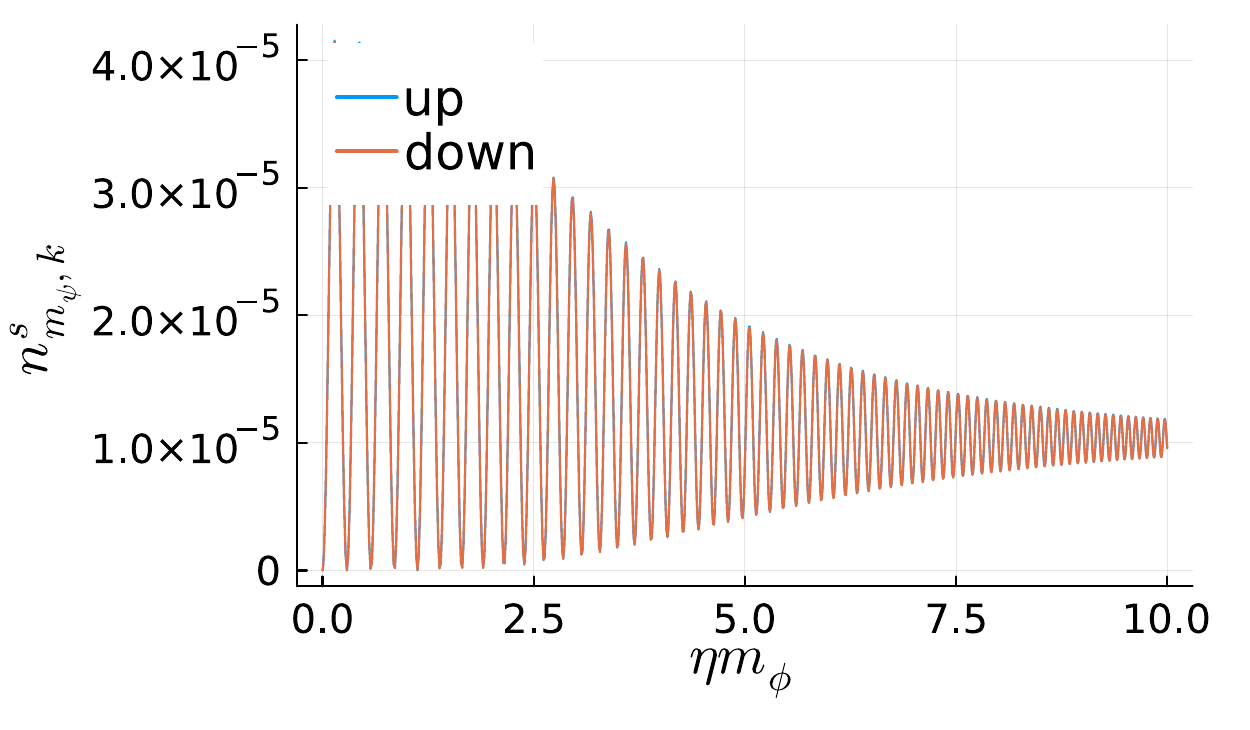}
      \subcaption{$m_{\psi}=4m_{\phi}$, $\bm{\beta=0}$, $k=10m_{\phi}$}
  \end{minipage}
  \caption{The time evolution of $n^s_{m_{\psi},k}$ with $k=10m_{\phi}$ and $m_{\psi}=4m_{\phi}$.}
  \label{figref:time_evolution}
\end{figure}

\begin{figure*}[htbp]
  \centering
  \begin{minipage}[h]{1\linewidth}
      \centering
      \includegraphics[width=0.9\linewidth]{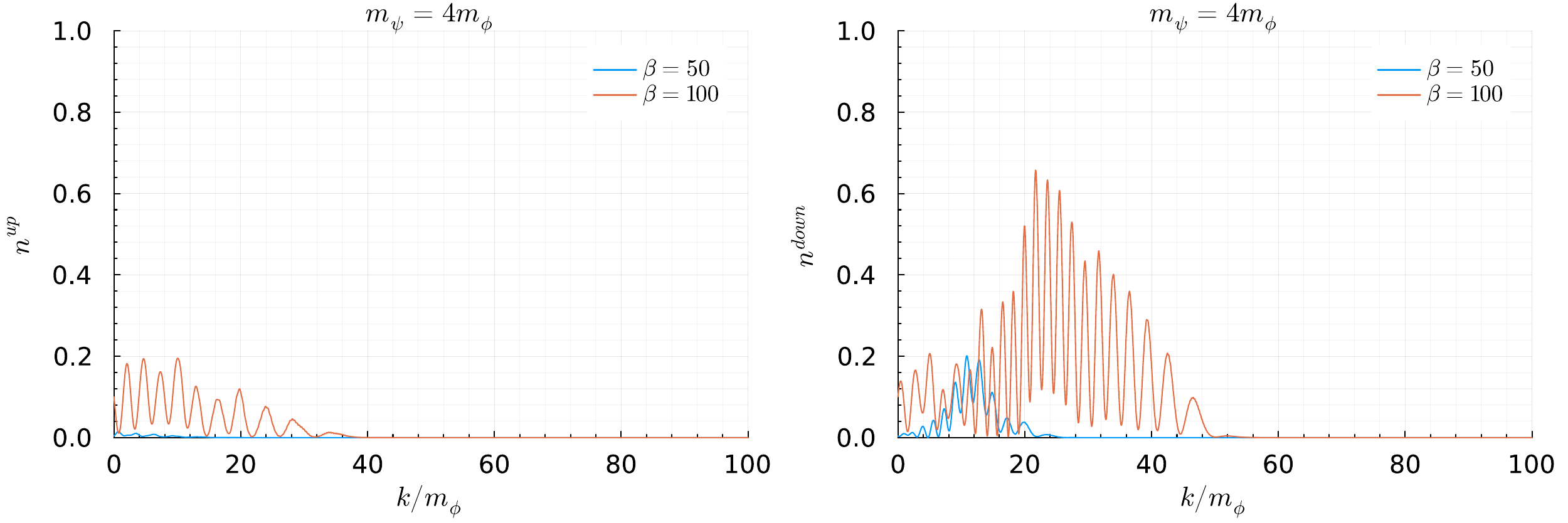}
  \end{minipage}
  \caption{The distribution of $n^s_{m_{\psi},k}$ as a function of the conformal momentum $k/m_{\phi}$ at $m_{\phi}\eta=20$ ($m_{\psi}=4m_{\phi}$).}
  \label{figure:numden}
\end{figure*}

\begin{figure*}[htbp]
  \centering
  \begin{minipage}[h]{0.9\linewidth}
      \centering
      \includegraphics[width=1\linewidth]{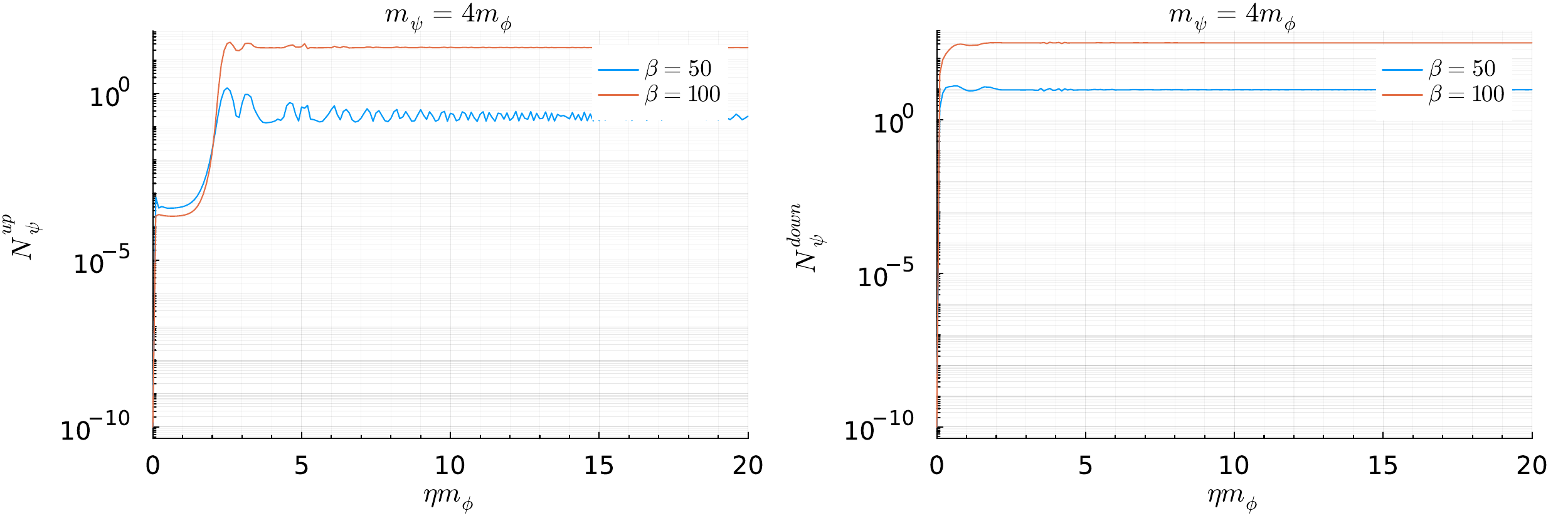}
  \end{minipage}
  \caption{The time evolution of $N^s_{\psi}$ ($m_{\psi}=4m_{\phi}$).}
  \label{figure:Number}
\end{figure*}

\begin{figure*}[htbp]
  \centering
  \begin{minipage}[h]{0.9\linewidth}
      \centering
      \includegraphics[width=1\linewidth]{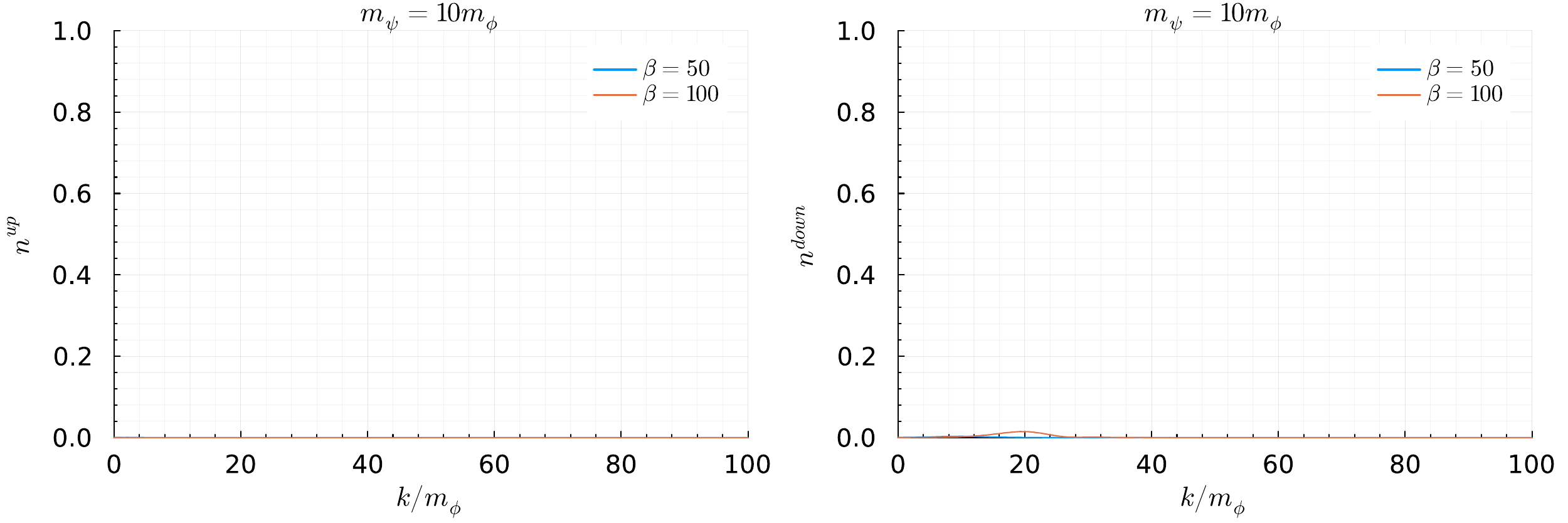}
  \end{minipage}
  \caption{The distribution of $n^s_{m_{\psi},k}$ as a function of the conformal momentum $k/m_{\phi}$ at $m_{\phi}\eta=20$($m_{\psi}=10m_{\phi}$).}
  \label{figure:m=10}
\end{figure*}

We evaluate the time evolution of the number density $n^s_{m_{\psi},k}$ for the helicity up and down particles with a certain conformal momentum $k$.
Fig.\ref{figref:time_evolution} clearly shows that the non-minimal coupling $\beta$ contributes to the non-thermal particle production.
Because of the interaction between the pseudoscalaron and the pseudovector current $\overline{\psi}i\gamma^5\gamma^{\mu}\psi$ in Eq.~\eqref{eq:eff_met_action}, the amount of the produced particles depend on the helicity.
In Fig.\ref{figure:numden}, we draw the distribution of $n^s_{m_{\psi},k}$ for $m_{\psi}=4m_{\phi}$ as a function of the conformal momentum $k/m_{\phi}$ for different $\beta$  at $\eta = 20m^{-1}_{\phi}$.
The upper limit of the conformal momentum for the non-thermal particle production is observed to be higher as $\beta$ increases.
Thus, a larger $|\beta|$ can lead to a higher conformal momentum excitation.
The extreme oscillations in Fig.\ref{figure:numden} are confirmed to be unaltered by the initial values of the inflaton and the solvers.
\\
After the end of of the particle production, the total number of the produced particles remains at a certain value~(Fig.\ref{figure:Number}).
In Figs.~\ref{figure:numden} and \ref{figure:Number}, different behavior is observed between up and down particles.
Both up and down particles are produced for $m_{\psi} = 4m_{\phi}$.
However, Fig.\ref{figure:m=10} shows that extremely heavy particles ($m_{\psi} = 10m_{\phi}$) are not produced.

\begin{figure*}[htbp]
  \begin{minipage}[h]{0.9\linewidth}
    \centering
    \includegraphics[width=1\linewidth]{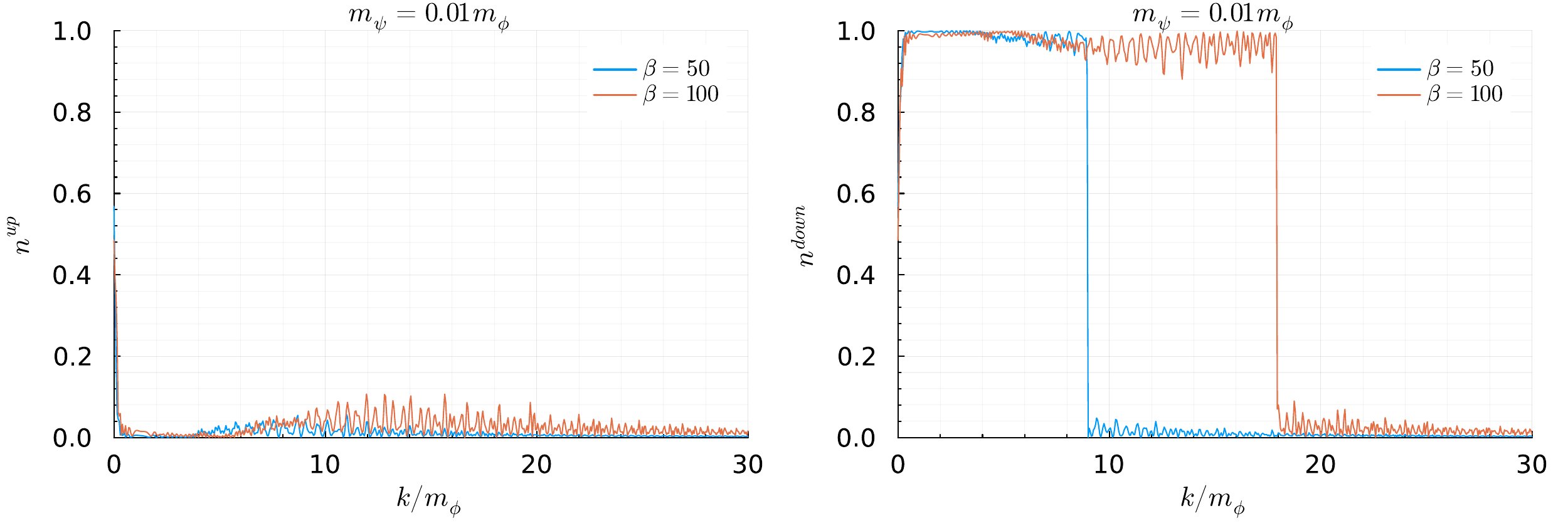}
  \end{minipage}
  \caption{The distribution of $n^s_{m_{\psi},k}$ as a function of the conformal momentum $k/m_{\phi}$ at $m_{\phi}\eta=100$($m_{\psi}=0.01m_{\phi}$).}
  \label{figure:numdenLight}
\end{figure*}

We also examine the number density distribution for lighter particles production.
Fig.\ref{figure:numdenLight} shows the number density distribution for $m_{\psi}=0.01m_{\phi}$ at $\eta = 100m^{-1}_{\phi}$.
It is observed that the higher conformal momentum particles can be produced with increasing $\beta$, as is mentioned in Sec.~4-1.
Compared with Fig.\ref{figure:numden}, the property in Tab.~\ref{table:masslessProduction} is almost confirmed for the lighter particles.

 From Eq.~\eqref{eq:initial_velocities}, the helicity is inverted when the sign of $f_0$ is reversed.
 A dominant contribution to $f_0$ comes from the second term in Eq.~\eqref{eq:inf_dyn1}.
 Thus, the up and down particles are exchanged and the aforementioned numerical results are nearly inverted for up and down with the sign of $\beta$ flipped.

\subsubsection{Intermediate-mass particle}
\begin{figure*}[htbp]
  \begin{minipage}[h]{0.9\linewidth}
    \centering
    \includegraphics[width=1\linewidth]{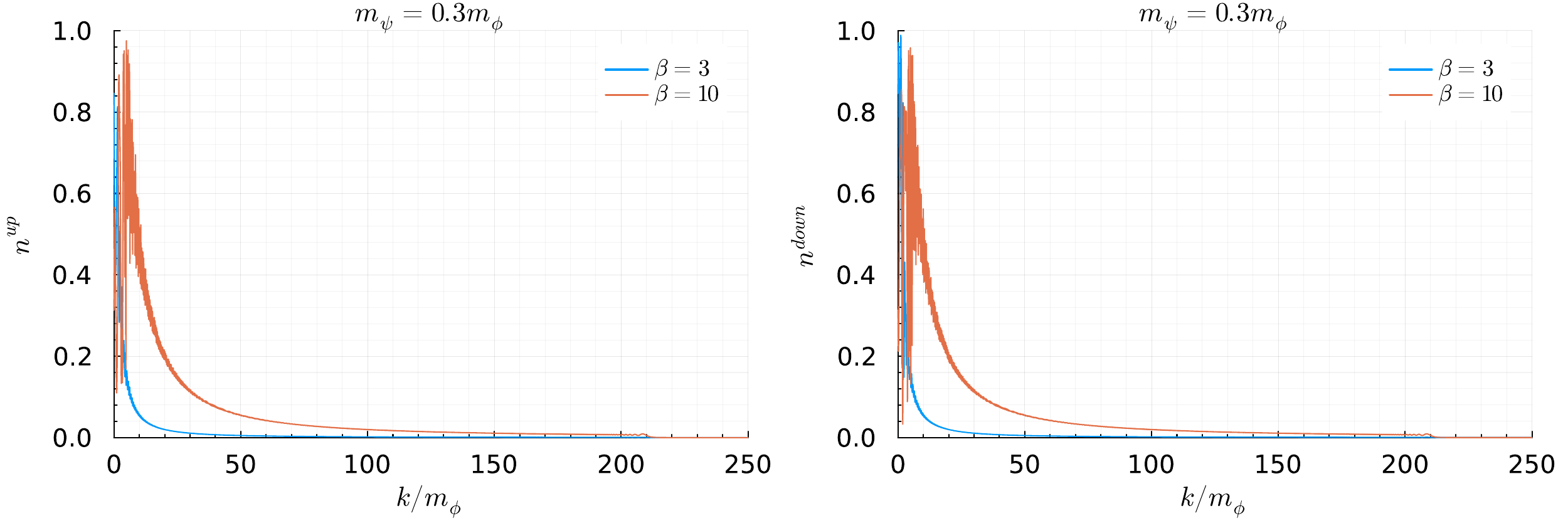}
  \end{minipage}
  \caption{The distribution of $n^s_{m_{\psi},k}$ as a function of the conformal momentum $k/m_{\phi}$ at $m_{\phi}\eta=110$($m_{\psi}=0.3m_{\phi}$).}
  \label{figure:numdenmedium}
\end{figure*}

For intermediate-mass particles, $m_{\psi}=0.1m_{\phi}\sim0.5m_{\phi}$, we observe an alternative property that does not appear in light and heavy particles. 
A fermionic field with a higher conformal momentum is gradually generated and the Boltzmann-type distribution is established.
According to the distribution of the number density (Fig.\ref{figure:numdenmedium}) for $m_{\psi}=0.3m_{\phi}$ at $\eta=110{m_{\phi}}^{-1}$, particles with greater conformal momentum are excited than that in the case of light and heavy particle. 

\begin{figure*}[htbp]
  \begin{minipage}[h]{0.9\linewidth}
    \centering
    \includegraphics[width=1\linewidth]{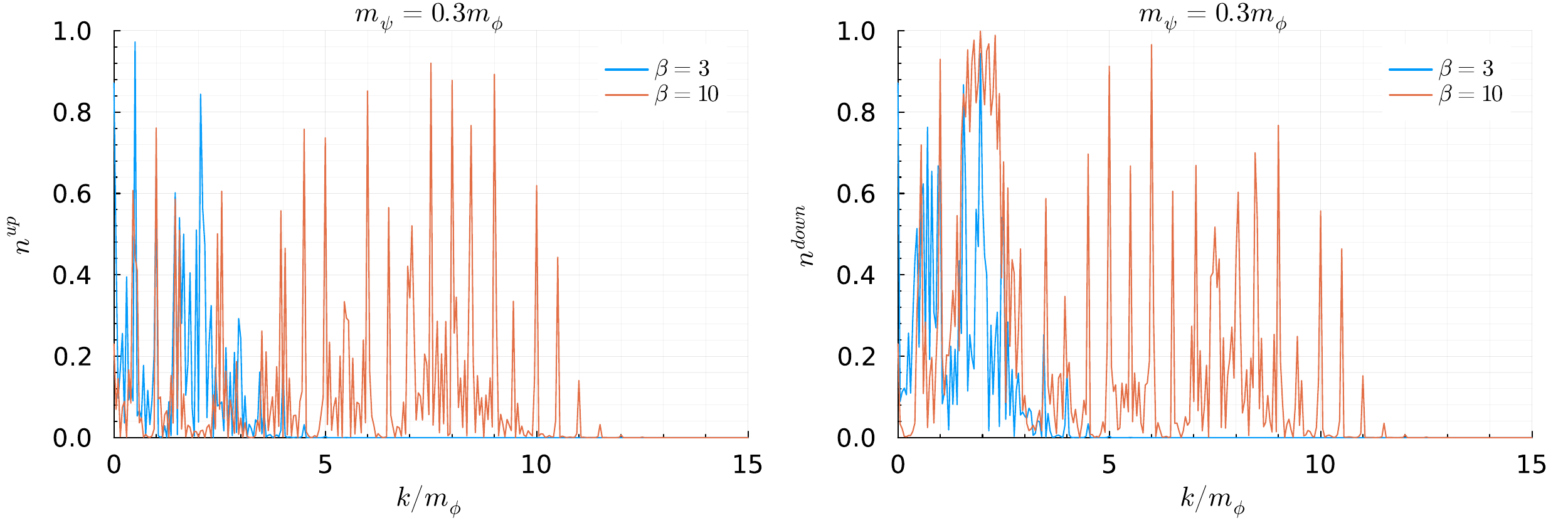}
  \end{minipage}
  \caption{The distribution of $n^s_{m_{\psi},k}$ as a function of the conformal momentum $k/m_{\phi}$ at $m_{\phi}\eta=110$($m_{\psi}=0.3m_{\phi}$) in non-expanding case.}
  \label{figure:nonexpanding}
\end{figure*}

We also evaluate the number density distribution in the non-expanding universe with a constant scale factor. 
The higher conformal momentum excitation is not observed in Fig.~\ref{figure:nonexpanding} for the non-expanding case.
It means that the higher conformal momentum excitation is due to the expansion of the universe.

\section{Cosmological consequences}
In this section, we apply the particle production to the reheating phenomena in the early universe.
In the standard history of the universe, the energy density of the matter field must exceed the energy density of the inflaton after the end of inflation.
Therefore, we examine if the condition $\rho_{\psi}\gg\rho_{\phi}$ can be achieved with the non-thermal and thermal particle production.

\subsection{Preheating} 
\begin{figure}[htbp]
  \centering
  \begin{minipage}[h]{1\linewidth}
      \centering
      \includegraphics[width=1\linewidth]{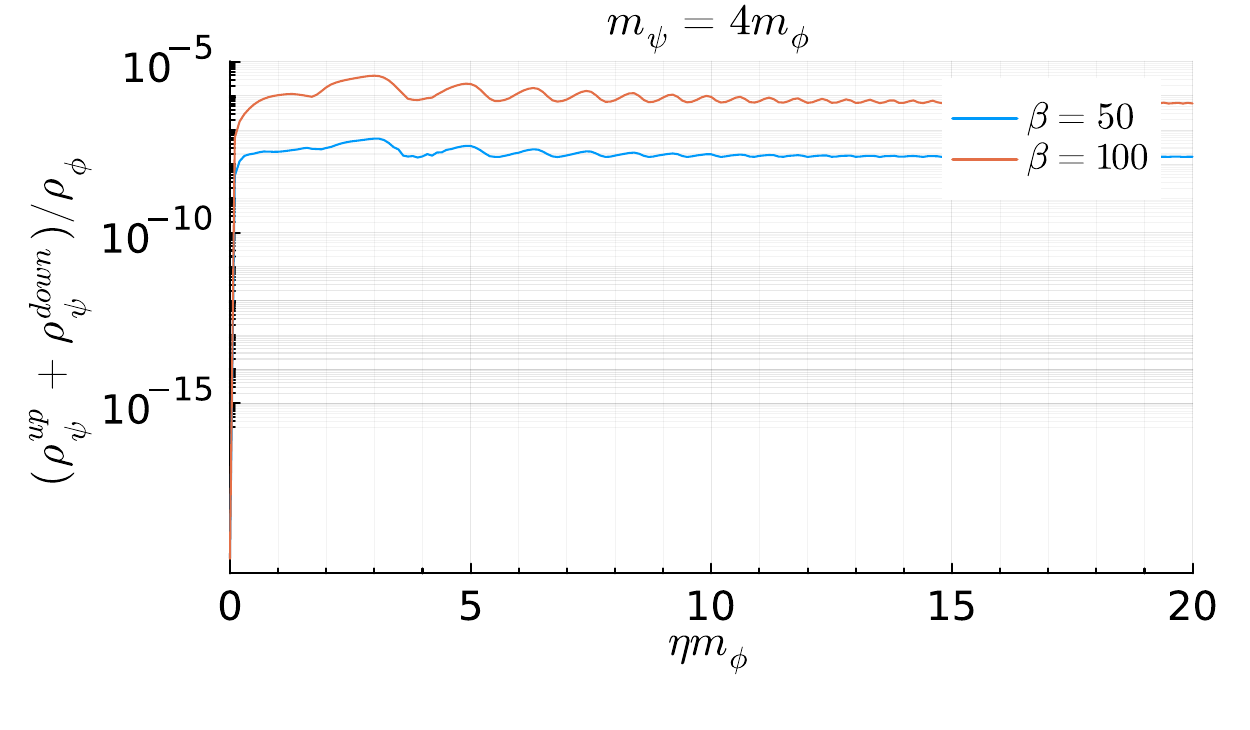}
  \end{minipage}
  \caption{The time evolution of the ratio of energy density $(\rho^{\text{up}}_{\psi} + \rho^{\text{down}}_{\psi})/\rho_{\phi}$ ($m_{\psi}=1m_{\phi}$).}
  \label{figure:eneden}
\end{figure}

The preheating is the thermal process of the universe due to the non-thermal particle production before the reheating. 
As we show in Chapter 4, a larger $|\beta|$ makes higher conformal momentum particles excited.
From Fig.\ref{figure:eneden}, ${\rho}_{\psi}/{\rho}_{\phi}$ grows as $\beta$ increases.
The condition ${\rho}_{\psi}/{\rho}_{\phi}\gg1$ can be achieved for sufficiently large $\beta$.

However, it is necessary to discuss thermalization due to the decay of produced particles into relativistic particles to define the reheating temperature. 
Thus, the results presented in this section only indicate that the energy can be sufficiently transferred from the inflaton to the matter.
Further analyses are required to estimate the reheating temperature.

\subsection{Reheating}
We adapt the perturvative calculations of the zero temperature quantum field theory to the fluctuation of $\phi$ after a long time from the end of inflation.
If the energy density of the inflaton $\rho_{\psi}$ transfers to that of relativistic matter $\rho_r$ through the decay with the decay rate $\Gamma$ and $\rho_r|_{\eta=0}=0$, the energy density, $\rho_r$, is estimated to be
\begin{align}
  \rho_r(t_r) = \frac{27}{25}\Gamma^2m^2_p,
\end{align}
at the moment for $\rho_{\phi} = \rho_r$.
From the Stefan-Boltzman law $\rho_r\propto T^4$, the reheating temparature $T_R$ is proportional to $\sqrt[]{\Gamma}$. 

In Einstein-Cartan pseudoscalaron model with non-minimal couplings to fermion, the decay of the $\phi$ into the fermions due to the interaction term 
\begin{align}
    f_{\mu}(\phi)A^{\mu} &\sim\frac{\sqrt[]{3}}{2\ \sqrt[]{2}m_p}\frac{1+\beta b}{\sqrt[]{1+b^2}}\overline{\psi}i\gamma^5\gamma^{\mu}\psi\partial_{\mu}\phi \notag\\ 
    &\equiv c_{\phi \overline{\psi}\psi}\overline{\psi}i\gamma^5\gamma^{\mu}\psi\partial_{\mu}\phi, \label{eq:int1}
\end{align}
dominates the particle production.
The decay rate $\Gamma_{\phi\overline{\psi}\psi}$ from this interaction is found to be
\begin{align}
  \Gamma_{\phi\overline{\psi}\psi} &= \frac{|c_{\phi \overline{\psi}\psi}|^2m_{\phi}m^2_{\psi}}{2\pi}\sqrt[]{1 - \Bigl(\frac{2m_{\psi}}{m_{\phi}}\Bigr)^2} \notag \\
                                   &=\frac{3m_{\phi}m^2_{\psi}(1 + b\beta)^2}{16\pi m^2_{p}(1+b^2)}\sqrt[]{1 - \Bigl(\frac{2m_{\psi}}{m_{\phi}}\Bigr)^2}. \label{eq:decayrate}
\end{align}
At $\beta=0$ the result \eqref{eq:decayrate} reproduces the one derived in the previous work \cite{Salvio:2022suk}.
Thus, even in thermal particle production, the effect of non-minimal coupling to fermion is significant for $|\beta b|\gg 1$. A reheating temperature is tuned by non-minimal coupling, $\beta$.
For $m_{\phi}\gg m_{\psi}$ and $|b\beta|\gg1$, reheating temperature is estimated as
\begin{align}
    T_R = \Bigl(\frac{30*27}{25\pi^2g^{\ast}}\Bigr)^{\frac{1}{4}}\sqrt{\frac{3}{16\pi}}\beta\Bigl( \frac{m_{\psi}}{m_{\phi}}\Bigr)^{1/2}m_{\phi},
\end{align}
where $k_B$ is the Boltzmann constant and $g^{\ast}$ shows the physical degree of freedom. 
The interaction of inflaton and the fermion vector current,
\begin{align}
    g_{\mu}(\phi)V^{\mu} \sim \frac{\sqrt[]{3}\alpha}{2\ \sqrt[]{2}m_p}\frac{b}{\sqrt[]{1+b^2}}\overline{\psi}i\gamma^{\mu}\psi\partial_{\mu}\phi,
\end{align}
has no contribution to the decay rate for $\phi\rightarrow\psi\overline{\psi}$, at the tree level.
Thus, $\alpha$ does not appear in Eq.~\eqref{eq:decayrate}.
The contribution from $\alpha$ dependent terms appear from the next to leading order.

\section{Summary and Discussion}
In this paper, we have investigated the non-thermal fermionic particle production in Einstein-Cartan gravity with modified Holst term and non-minimal couplings to fermion.

In EC gravity,
the only imposed condition is metric compatiblity, $\nabla_{\mu}g_{\nu\rho}=0$.
Thus, the antisymmetric part of the affine connection ${T^\mu}_{\nu\rho}\equiv{\Gamma^{\mu}}_{\nu\rho}-{\Gamma^{\mu}}_{\rho\nu}$(i.e., torsion) can have a non-vanishing value.
The existence of the torsion introduces invariant scalar components which do not exist in GR.
One of these quantities is the Holst term $\epsilon R=\epsilon^{\mu\nu\rho\sigma}R_{\mu\nu\rho\sigma}/\sqrt[]{-g}$.
The derivative of the affine connection in this term introduces a dynamical degree of freedom through the EoM.
Following the auxiliary field method, the EoM for the affine connection becomes an algebraic equation for the torsion.
After solving the EoM of the torsion, one can get the effective metric action with the dynamical pseudoscalaron $\phi$ \cite{Pradisi:2022nmh, Salvio:2022suk,DiMarco:2023ncs,He:2024wqv,Barker:2024dhb}.
The potential energy of the pseudoscalaron can induce inflation in EC gravity.
In this paper, we utilize the $(\epsilon R)^2$ model, which is consistent with the CMB observations.

The coupling between the matter fields and the affine connection in the original action yields 
an interaction between the matter fields and the inflaton in the effective metric action.
In this study, we have employed a theory with non-minimal couplings to fermion where two parameters denoted as $\alpha$ and $\beta$ are introduced \cite{Freidel:2005sn}.
This extension introduces interactions between the inflaton and the fermionic field $\psi$ described in Eq.~\eqref{eq:eff_met_action}.
Since the inflaton $\phi$ has a large value after the end of inflation, 
the interactions between the $\phi$ and the $\psi$ destabilize the vacuum.
Through the instability, the non-thermal particle production occurs.
Since the parameter $\alpha$ must be zero to satisfy the condition that $\mathcal{C}\overline{\psi}^T$ represents a field where particles and antiparticles are exchanged, 
we have shown that the non-minimal coupling $\beta$ in Eq.~\eqref{eq:eff_met_action} contributes to the non-thermal particle production after the end of inflation.
In this paper, only $\phi\overline{\psi}\psi$ terms in Eq.~\eqref{eq:decomposition} that introduce linear term with respect to $\psi$ to the EoM of $\psi$ is focused.

We have examined how many particles are produced due to the existence of $\beta$ through numerical calculations and observed several properties.
First, a larger value of $\beta$ leads to the excitation of particles with higher conformal momentum, and the excitation persists over time.
It is also observed that fermions much heavier than the inflaton are hardly excited.
Second, there is generally a difference in the amount of produced particles between helicity up and down particles.
It should be noted that the produced numbers of particles and antiparticles are identical, and the total spin of the universe is conserved.
Third, for the lighter mass fermion, $m_{\psi} \lessapprox 0.1m_{\phi}$, the difference in the number of created particles between helicities becomes more pronounced.
This property is also suggested analytically at the massless limit.
If the initial value of $\frac{d\phi}{d\eta}$ is small enough, light particles are not excited non-thermally.
Eventually, for the intermediate-mass fermion $0.1m_{\phi} \lessapprox m_{\psi} \lessapprox 0.5m_{\phi}$,
 the number density
  is exponentially supressed for a higher conformal momentum like a Boltzmann distribution. Therefore, particles with higher conformal momentum can be excited than for heavier and lighter fermions.
This property is not observed in a non-expanding universe.

We have applied the particle production to the reheating and the preheating of the universe.
From the consistency with the CMB observations, we set a model paremeter $b\sim-320$ in our analysis.
In the preheating era, sufficiently large values of $\beta$ have the potential to make the energy density of $\psi$ dominant over that of the inflaton.
In the reheating era, we derive the formula of the reheating temperature $T_{R, \beta}$ with non-minimal coupling.
About $|\beta b|$ times larger reheating temperature is predicted than that for the minimal coupling \cite{Salvio:2022suk}.
Thus, the contribution of $\beta$ is important in both eras.

Future research directions include further development of analytical and numerical discussions and applications to phenomenology.
Due to the complexity of the function of the inflaton coupling to fermion as given in Eq.~\eqref{eq:inf_dyn1}, 
the dynamics of fermions during inflation and the backreaction to the evolution of the universe during particle production are not considered. 
To discuss a realistic universe, consideration of these two aspects can not be avoidable. 
Moreover, analytical solutions in some limits are necessary to verify the numerical results.
Potential applications of our research include the production of the dark matter and the matter-antimatter asymmetry. 
The particle production investigated in our research is induced by gravitational effect.
Heavy fermions that could be dark matter may be produced after the end of inflation. 
Our asymmetric helicity production has a possibility to describe the matter-antimatter asymmetry through particle production of fermions with lepton numbers \cite{Adshead:2015kza,Adshead:2015jza}.
Our analysis can be applied to Majorana fermion which directly induces lepton number asymmetry \cite{Fukugita:1986hr}.

\begin{acknowledgments}
The authors would like to thank H.~Sakamoto, M.~Alwan, M.~Taniguchi and S.~Takahashi for valuable discussion.
\end{acknowledgments}

\newpage
\bibliography{main}

\end{document}